\documentclass[11pt,a4paper]{article}
\usepackage{amsfonts}
\usepackage{amsmath,amssymb}
\usepackage{epsfig,graphicx}

\input{tcilatex}

\begin{document}

\begin{center}
{\LARGE \ Spontaneous Lorentz Violation via QED with Non-Exact Gauge
Invariance }

\bigskip

\textbf{J.L.~Chkareuli}, \ \textbf{Z. Kepuladze\ and G. Tatishvili}

\textit{E. Andronikashvili} \textit{Institute of Physics and }

\textit{I. Chavchavadze State University, 0177 Tbilisi, Georgia\ \vspace{0pt}%
\\[0pt]
}

\bigskip \bigskip \bigskip \bigskip \bigskip \bigskip
\bigskip\bigskip\bigskip

\textbf{Abstract}

\bigskip
\end{center}

We reconsider an alternative theory of the QED with the photon as a massless
vector Nambu-Goldstone boson and show that the underlying spontaneous
Lorentz violation caused by the vector field vacuum expectation value,
while\ being\ superficial in gauge invariant theory, becomes physically
significant in the QED with a tiny gauge non-invariance. This leads, through
special dispersion relations appearing for charged fermions, to a new class
of phenomena which could be of distinctive observational interest in
particle physics and astrophysics. They include a significant change in the
GZK cutoff for UHE cosmic-ray nucleons, stability of high-energy pions and $%
W $ bosons, modification of nucleon beta decays, and some others.

\thispagestyle{empty}\newpage

\section{Introduction}

Relativistic or Lorentz invariance, while still perfectly fits nature
observed, might be broken at high energies thus manifesting itself in some
new phenomena being presently hidden. This has attracted considerable
attention in recent years as an interesting phenomenological possibility
appearing in direct Lorentz non-invariant extensions of QED and Standard
Model (SM) [1,2,3]. These extensions may in a certain measure be motivated
[4] by a string theory where a spontaneous Lorentz violation can occur when
a theory has a non-perturbative vacuum that leads to tensor-valued fields
acquiring non-zero vacuum expectation values (vevs). The vevs are
effectively a set of coupling constants so that interactions with these
coefficients have preferred spacetime directions. The full SM extension
(SME) [2] is then defined as the effective gauge invariant field theory
obtained when all such Lorentz violating vevs are contracted term by term
with SM (and gravitational) fields. However, without a completely viable
string theory, it is not possible to assign definite numerical values to
these coefficients. Therefore, one has in this sense a pure phenomenological
approach treating the above arbitrary coefficients as quantities to be
bounded in experiments as if they would simply appear due to explicit
Lorentz violation. Actually, there is nothing in the SME by itself that
requires that these Lorentz-violation coefficients emerge due to a process
of spontaneous Lorentz violation -- neither the corresponding massless
vector (tensor) Nambu-Goldstone (NG) bosons are required to be generated as
extra physical states in the Standard Model, nor (especially) do these
bosons have to be associated with photons or any other gauge fields of the
SM.

On the other hand, however, Lorentz invariance seems to play a special role
with respect to the observed internal local symmetries. The old idea [5]
that spontaneous Lorentz invariance violation (SLIV) may lead to an
alternative theory of the QED, with the photon as a massless vector NG
boson, still remains extremely attractive in numerous theoretical contexts
[6]. In the present paper we will follow this genuine SLIV pattern causing
dynamical generation of physical gauge fields rather than the above SME
approach providing a general phenomenological framework for Lorentz
violation. Specifically, we focused here on the question of how this type of
the SLIV (triggered by the vector field vev), while being superficial in
gauge invariant theory, may become physically significant in the QED with a
tiny gauge non-invariance. Notably, in contrast to the gauge invariant SME,
physical Lorentz violation in the genuine SLIV model can only occur if this
gauge invariance is broken. We find that such a possibility may appear when
the theory is extended to include the higher dimension operators in the
matter and vector fields involved. Remarkably at the same time, a special
form of physical Lorentz violation arising in the minimal dimension-five QED
model happens to be one of many possible breaking patterns emerging in a
general SME expansion [2]. This means in turn that our model is expected to
be rather definite in its experimental predictions.

Before proceeding, we briefly recall some of generic ingredients of this
SLIV\ approach which started long ago [5] in terms of models based on the
four-fermion (current $\times $ current) interaction, where the Goldstonic
gauge field may appear as a composite fermion-antifermion state.
Unfortunately, owing to the lack of an initial gauge invariance in such
models and the composite nature of the NG modes that appear, it is hard to
explicitly demonstrate that these modes really form together a massless
vector boson as a gauge field candidate. Actually, one must make a precise
tuning of parameters, including a cancellation between terms of different
orders in the $1/N$ expansion (where $N$ is the number of fermion species
involved), in order to achieve the massless photon case (see, for example,
the last paper in [5]). Rather, there are in general three separate massless
NG modes, two of which may mimic the transverse photon polarizations, while
the third one must be appropriately suppressed.

In this connection, the more instructive laboratory for SLIV consideration
proves to be some simple class of the QED type models [7,8] having from the
outset a gauge invariant form whereas the spontaneous Lorentz violation is
realized through the non-linear dynamical constraint 
\begin{equation}
A^{2}=n^{2}M^{2}\qquad (A^{2}\equiv A_{\mu }A^{\mu },\quad n^{2}\equiv
n_{\mu }n^{\mu })  \label{con}
\end{equation}%
(where $n_{\mu }$ is a properly oriented unit Lorentz vector, $n^{2}=\pm 1$,
while $M$ is a proposed SLIV scale) imposed on the starting vector field $%
A_{\mu }$ This constraint means in essence that the vector field $A_{\mu }$
develops some constant background value $<A_{\mu }(x)>$ $=(n_{\mu }/n^{2})M$%
\ and Lorentz symmetry $SO(1,3)$ breaks down to $SO(3)$ or $SO(1,2)$
depending on the time-like ($n^{2}>0$) or space-like ($n^{2}<0$) SLIV. This
violation provides in fact the genuine Goldstonic nature of QED, as could
easily be seen from an appropriate $A_{\mu }$ field parametrization, 
\begin{equation}
A_{\mu }=a_{\mu }+\frac{n_{\mu }}{n^{2}}(M^{2}-n^{2}a_{\nu }^{2})^{\frac{1}{2%
}},\text{ \ }n_{\mu }a^{\mu }=0\text{\ \ }  \label{par}
\end{equation}%
where the pure Goldstone modes $a_{\mu }$ are associated with photon, while
an effective Higgs mode, or the $A_{\mu }$ field component in the vacuum
direction, is given by the square root in (\ref{par}).

Actually, to appreciate the possible origin for the supplementary condition\
(1) one might consider the inclusion of a \textquotedblleft
standard\textquotedblright\ quartic vector field potential%
\begin{equation}
U(A)=-\frac{m_{A}^{2}}{2}A^{2}+\frac{\lambda _{A}}{4}(A^{2})^{2}  \label{pol}
\end{equation}%
in the conventional QED Lagrangian, as could be motivated [4] to some extent
from a string theory. This potential inevitably causes the spontaneous
violation of Lorentz symmetry in a conventional way, much as an internal
symmetry violation is caused in a linear $\sigma $ model for pions [9]. As a
result, one has a massive Higgs mode (with mass $\sqrt{2}m_{A}$) together
with massless Goldstone modes associated with the photon. Furthermore, just
as in the pion model, one can go from the linear model for the SLIV to the
non-linear one by taking the limit $\lambda _{A}\rightarrow \infty ,$ $%
m_{A}^{2}\rightarrow \infty $ (while keeping the ratio $m_{A}^{2}/\lambda
_{A}$ finite). This immediately leads to the constraint (1) for the vector
potential $A_{\mu }$ with $n^{2}M^{2}=m_{A}^{2}/\lambda _{A}$, as appears
from the validity of its equation of motion. Note that a correspondence with
the non-linear $\sigma $ model for pions may be somewhat suggestive, in view
of the fact that pions are the only presently known Goldstones and their
theory, chiral dynamics [9], is given by the non-linearly realized chiral $%
SU(2)\times SU(2)$ symmetry rather than by an ordinary linear $\sigma $
model.

The point is, however, that, in sharp contrast to the non-linear $\sigma $
model for pions, the non-linear QED theory, due to the starting gauge
invariance involved, ensures that all the physical Lorentz violating effects
are proved to be non-observable: the SLIV condition (1) is simply reduced to
a possible gauge choice for the vector field $A_{\mu }$, while the $S$%
-matrix remains unaltered under such a gauge convention [7]. Really, this
non-linear QED contains a plethora of particular Lorentz and $CPT$ violating
couplings when expressed in terms of the pure Goldstonic photon modes $%
a_{\mu }$ (2). However, contributions of these couplings to all physical
processes involved are\ proved to be strictly cancelled, as was explicitly
demonstrated in tree [7] and one-loop [8] approximations.

So, whereas it seems that the photon could very likely have a true
Goldstonic nature, the most fundamental question of whether the physical
Lorentz violation takes place, that only might uniquely point toward such a
possibility, is still an open question. Recall that we do not touch here on
direct Lorentz non-invariant extensions of QED or Standard Model when some
non-covariant vector and/or matter field combinations (bilinears, trilinears
etc.) are explicitly introduced in theory [1,2,3]. Instead, we are looking
for a pure SLIV framework which, hand in hand with a photon appearing as a
proper vector NG boson, could produce observable Lorentz violating effects.

In this context, for physical Lorentz violation to occur, an internal gauge
symmetry in the theory considered should be explicitly broken rather than
exact or spontaneously violated\footnote{%
Physical SLIV effects appear to be entirely cancelled in a general Abelian
theory as well (see the last paper in Ref.[8]), particularly, in the case
when the internal $U(1)$ charge symmetry is spontaneously broken hand in
hand with Lorentz invariance. As a result, the massless photon being first
generated by the Lorentz violation then becomes massive due to the standard
Higgs mechanism, while the SLIV condition (1) in itself remains as a pure
gauge choice.}. We propose that such a tiny gauge non-invariance might
appear at very short distances through some higher dimension operators
stemming from the gravity-influenced area. If so, physical SLIV effects
would be seen in terms of powers of the ratio $M/\mathcal{M}$, where the
scale $\mathcal{M}$ might be related to the Planck mass $M_{P}$, as would
appear in some string theory scenarios, or a certain compactification scale.
Notably enough, if one has such internal gauge symmetry breaking in an
ordinary Lorentz invariant theory, this breaking appears to be vanishingly
small at low energies, being properly suppressed by the scale $\mathcal{M}$.
However, the spontaneous Lorentz violation would render it physically
significant: the higher the scale $M$, the greater the SLIV\ effects
observed. Remarkably, the gauge non-invariance proposed cannot generate the
photon mass since photons appear in the theory as the vector NG bosons and,
therefore, their masslessness is guarantied by the SLIV. An absence of
longitudinal photons in the theory together with strict conservation of the
Noether fermion current involved provides, on the other hand, conservation
of electric charge as well\footnote{%
Note, at the same time, that if electric charge non-conservation would
occur, this violation would appear, as we see in Sec.2, well below the
presently existing bounds \cite{pdg}.}.

To put \ all this another way, note that gauge invariance in the Goldstonic
QED appears in essence as a necessary condition for the starting vector
field $A_{\mu }$ not to be superfluously restricted in degrees of freedom,
apart from the SLIV constraint (1) due to which the true vacuum in the
theory is chosen [11]. For any extra restriction(s) imposed on the vector
field, it would be impossible to set the required initial conditions in the
appropriate Cauchy problem and, in quantum theory, to choose self-consistent
equal-time commutation relations [12]. From this point of view, the only
possible theory compatible with the SLIV condition (1) appears to be\ just a
conventional gauge invariant QED. One may expect, however, that quantum
gravity could in general hinder the setting of the required initial
conditions at extra-small distances thus admitting superfluous restriction
of the starting vector field $A_{\mu }$. This eventually, through some
high-order operators, would manifest itself in violation of the above gauge
invariance that in turn might bring the spontaneous Lorentz violation to low
energies. We suggest here such a type of model (Sec.2) and explore some of
its immediate physical and astrophysical consequences (Sec.3). Our
conclusions are given in the final Sec.4.

\section{The Model}

Before proceeding to the extension of a conventional QED to the higher
dimension couplings included, we are reminded that gauge invariance in a
standard quantum electrodynamics is not necessarily postulated for the
photon-fermion interaction - it appears on its own if, apart from
relativistic invariance, the restrictions related with the conservation of
parity, charge-conjugation symmetry and number of fermions are also imposed
in the Lagrangian. Actually, one uses gauge invariance only if one
constructs the photon kinetic term to have an ordinary $F_{\mu \nu }F^{\mu
\nu }$ form since this is necessary in order that the Hamiltonian be bounded
below\footnote{%
Note also that a general photon kinetic term gives rise to ghosts in the
propagator and, specifically in the Lorentz-violating QED type theory, to a
domain wall solution for the vector potential $A_{\mu }$ that might lead to
a wall-dominated early Universe and its immediate collapse \cite{dgs}.}.
Similarly, analogous restrictions for photon-fermion couplings of higher
dimensions generally allow only for a few new ones (for each order in the
theory's inverse scale $1/\mathcal{M}$) which appear to possess, however,
some approximate gauge invariance rather than an exact one as one has in a
conventional QED with dimensionless coupling constants. In this connection
the most transparent situation arises in a minimal QED extension to
dimension-five couplings which we consider here in detail. Since this
extension, apart from photon-fermion interaction terms, will necessarily
include the free fermion bilinear of type $(1/\mathcal{M)}\partial _{\mu }%
\overline{\psi }\partial ^{\mu }\psi $ one could hold to the idea that free
fermions would generally be described by some combined Dirac-Klein-Gordon
equation rather than the pure Dirac equation that might be hidden at low
energies. However, due to spontaneous Lorentz violation this "fermion-boson
complementarity" could become significant providing a somewhat natural model
for a tiny gauge non-invariance in the QED when the electromagnetic
interaction is "switched on". As a result, the SLIV, having been\
superficial in gauge invariant theory, becomes in fact physically observable
through a certain dispersion relation which automatically appears for
charged fermions. This is in contrast to the direct Lorentz violation models
[2,3] where some modified dispersion relations for the photon and/or matter
particles involved are\ in essence specially postulated.

One can start with a free Lagrangian for some massive charged fermion $\psi $
in the form 
\begin{equation}
L(\psi )=\overline{\psi }(i\gamma _{\mu }\partial ^{\mu }-m_{0})\psi +\frac{1%
}{\mathcal{M}}\partial _{\mu }\overline{\psi }\cdot \partial ^{\mu }\psi 
\label{free}
\end{equation}%
which contains, apart from a true fermionic kinetic term, some "bosonic"
type kinetic term as well. As is clear from this Lagrangian, the fermion
dispersion relation will be a little changed so that for its four-momentum $%
p_{\mu }$ squared one has 
\begin{equation}
p^{2}=(m_{0}-p^{2}/\mathcal{M})^{2}=m_{0}^{2}(1-2m_{0}/\mathcal{M}+\cdot
\cdot \cdot )
\end{equation}%
which leads to a tiny mass shift for fermion which, of course, is of no
experimental interest. Let us now turn on all possible interaction terms
which, under the foregoing discrete and global symmetry restrictions taken,
amount to the gauge type "minimal" interactions of fermion with vector field
(given by the standard replacement $\partial ^{\mu }\rightarrow \partial
^{\mu }+ieA^{\mu }$) through both of kinetic terms involved in the
Lagrangian $L$ (\ref{free}). In this connection, there might appear the
question of whether the \textquotedblleft fermionic\textquotedblright\ and
\textquotedblleft bosonic\textquotedblright\ type couplings of the $\psi $
field with the vector field $A^{\mu }$ have the same coupling constant $e$.
If so, the total Lagrangian with the above "minimal" interaction included,
while being non-renormalizible, will still be left gauge invariant. However,
generally, these coupling constants are different, which means that the
Lagrangian is no more gauge invariant as soon as one takes into account the
small "bosonic" type kinetic term in (\ref{free}) being suppressed by the
scale $\mathcal{M}$. This is just a type of gauge non-invariance that
underlies our model leading eventually to physical Lorentz violation. So,
the initially Lorentz invariant theory for fermion-vector field
interactions, which possesses a slightly broken gauge invariance, is given
by the general Lagrangian\footnote{%
For simlicity we have not included into the Lagrangian $L\left( A,\psi
\right) $ the anomalous magnetic moment type coupling $\frac{e^{\prime
\prime }}{\mathcal{M}}F^{\mu \nu }\overline{\psi }\sigma _{\mu \nu }\psi $
which is gauge invariant on its own and appears inessential for what
follows. Another simplification is that we have omitted an independent
"sea-gull" type coupling $\frac{e^{\prime \prime \prime }}{\mathcal{M}}%
A_{\mu }^{2}\overline{\psi }\psi $ in the Lagrangian (a term like that is
already contained in its "bosonic" part), since such a coupling due to the
SLIV condition (1) is simply reduced to some inessential correction to the
fermion mass term. All things considered, the Lagrangian $L\left( A,\psi
\right) $ gives in fact the most general extension of QED in $\frac{1}{%
\mathcal{M}}$ order, taken under the Lorentz and extra discrete and global
symmetry restrictions discussed above.} 
\begin{equation}
L\left( A,\psi \right) =-\frac{1}{4}F_{\mu \nu }F^{\mu \nu }+\overline{\psi }%
[i\gamma _{\mu }D^{\mu }-m_{0}]\psi +\frac{1}{\mathcal{M}}D_{\mu }^{\prime
\ast }\overline{\psi }\cdot D^{\prime \mu }\psi   \label{lagr}
\end{equation}%
containing, apart from the "true fermionic" terms with covariant derivative $%
D^{\mu }=\partial ^{\mu }+ieA^{\mu }$, the "bosonic" type terms as well $\ $%
with $D^{\prime \mu }=\partial ^{\mu }+ie^{\prime }A^{\mu }$, either taken
with independent charges $e$ and $e^{\prime }$, respectively. Remarkably,
despite the fact that both the \textquotedblleft
fermionic\textquotedblright\ and \textquotedblleft
bosonic\textquotedblright\ parts of the Lagrangian (\ref{lagr}) are
individually gauge invariant, gauge invariance is in fact broken when they
are taken together. As a result, though this Lagrangian practically (i.e.
neglecting the last term in (\ref{lagr})) does not differ from a
conventional QED Lagrangian, provided that the vector field $A_{\mu }$ is
associated with a photon, a drastic difference appears when this field
develops a vev and the SLIV occurs.

Actually, putting the SLIV parameterization (\ref{par}) into our basic
Lagrangian (\ref{lagr}) one comes to the truly Goldstonic model for the QED.
This model contains, among other terms, the inappropriately large (while
false) Lorentz violating fermion bilinear $-eM\overline{\psi }(n_{\mu
}\gamma ^{\mu }/n^{2})\psi $, which appears when the effective Higgs field
expansion (as is given in the parametrization (\ref{par})) in true Goldstone
modes $a_{\mu }$ is applied to the fermion current interaction term $-%
\overline{\psi }\gamma _{\mu }A^{\mu }\psi $ in the "fermionic" part of the
Lagrangian $L\left( A,\psi \right) $. However, due to local invariance of
this part, this bilinear term can be gauged away by making an appropriate
redefinition of the fermion field $\psi \rightarrow e^{-ie\omega (x)}\psi $
with a gauge function $\omega (x)$ linear in coordinates, $\omega (x)=$ $%
(n_{\mu }x^{\mu }/n^{2})M$. Meanwhile, the small "bosonic" part being gauge
non-invariant is appropriately changed under this redefinition. So, one
eventually arrives at the essentially non-linear SLIV Lagrangian for
photon-fermion interaction with the significantly modified fermion bilinear
terms 
\begin{eqnarray}
\mathcal{L}\left( a_{\mu },\psi \right)  &=&L\text{ {\Large (}}A_{\mu
}\rightarrow a_{\mu }+n_{\mu }(a^{2}/2M+\cdot \cdot \cdot ),\psi \text{%
{\Large )}}+  \label{NL} \\
&&-i\Delta e\frac{M}{\mathcal{M}}\frac{n_{\mu }}{n^{2}}\overline{\psi }%
\overleftrightarrow{\partial ^{\mu }}\psi +(\Delta e)^{2}n^{2}\frac{M^{2}}{%
\mathcal{M}}\overline{\psi }\psi   \notag
\end{eqnarray}%
where we have explicitly indicated that the vector field $A_{\mu }$ in the
starting Lagrangian $L$ (\ref{lagr}) is replaced by the pure Goldstone field 
$a_{\mu }$ associated with the photon (appearing in the gauge $n_{\mu
}a^{\mu }=0$ ) plus the effective Higgs field expansion in (\ref{par}). We
also retained the notation $\psi $ for the redefined fermion field and
denoted, as usually, $\overline{\psi }\overleftrightarrow{\partial ^{\mu }}%
\psi =\overline{\psi }(\partial ^{\mu }\psi )-(\partial ^{\mu }\overline{%
\psi })\psi $. Note that the extra fermion bilinear terms\footnote{%
Notably, from a general SME point of view one could say that just this form
of physical Lorentz violation known as "$e$-term" breaking [2] appears to
dominate, due to the genuine SLIV pattern considered, over many other
Lorentz breaking terms emerging in the SME.} given in the second line in (%
\ref{NL}) are produced just due to the gauge invariance breaking that is
determined by the electromagnetic charge difference $\Delta e=e^{\prime }-e$
in the starting Lagrangian $L$ (\ref{lagr}). As a result, there appears the
entirely new, SLIV\ inspired, dispersion relation for a charged fermion
(taken with 4-momentum $p_{\mu }$) of the type%
\begin{equation}
p_{\mu }^{2}\cong \lbrack m+2\delta (p_{\mu }n^{\mu }/n^{2})]^{2},\text{ \ \ 
}m=m_{0}-\delta ^{2}n^{2}\mathcal{M}  \label{dr}
\end{equation}%
given to an accuracy of $O(m^{2}/\mathcal{M}^{2})$. Here $\delta $ stands
for the small characteristic, positive or negative, parameter $\delta
=(\Delta e)M/\mathcal{M}$ of the physical Lorentz violation that reflects
the joint effect as given, from the one hand, by the SLIV scale $M$ and,
from the other, by the charge difference $\Delta e$ being a measure of an
internal gauge non-invariance. Notably, the space-time by itself still
possesses Lorentz invariance, however, fermions with the SLIV contributing
into their total mass $m=m_{0}-\delta ^{2}n^{2}\mathcal{M}$ propagate and
interact in it in the Lorentz non-covariant way\footnote{%
Note also that the fermion dispersion relation (\ref{dr}) is substantially
different from the dispersion relations extensively used before [3] where
just the mass squared happened to be shifted in the preferred frame rather
than the mass by itself as in Eq.(\ref{dr}).}. At the same time, the photon
dispersion relation\ is retained in order $1/\mathcal{M}$ considered%
\footnote{%
One must, of course, expect that non-gauge invariant photon kinetic terms,
changing its dispersion relation, are also generated through radiative
corrections. But these terms are down by high orders in $1/\mathcal{M}$
relative to the basic $F_{\mu \nu }^{2}$ term taken, and, therefore, can be
neglected.}.

Let us now try to estimate a possible scale of Lorentz violation and a
numerical value of the parameter $\delta $ being in essence the only measure
of the physical Lorentz violation in our model. Some estimation could follow
from the naturalness requirement that the free fermion mass presented in the
Goldstonic QED Lagrangian (\ref{NL}) and, specifically the mass of the
lightest charged fermion which is the electron mass, should not be
significantly disturbed by the Lorentz violation. Otherwise possible fine
tuning between the SLIV contribution to this mass and its starting value
would become necessary. Proposing the SLIV contributed total electron mass $%
m_{e}$ to remain of the same order as the starting mass $m_{0e}$, one comes
from (\ref{dr}) to the inequality $\delta ^{2}\mathcal{M}\lesssim m_{e}$.
Remarkably, the characteristic parameter $\delta $ depends on neither the
SLIV scale $M$ nor on the charge difference $\Delta e$ individually but on
their product only, and, for the above "stability condition" against the
SLIV contribution to the electron mass, it is generally given by the range
of values%
\begin{equation}
\delta =(\Delta e)M/\mathcal{M},\text{ \ \ }\left\vert \delta \right\vert
\lesssim \overline{\delta }\equiv \sqrt{m_{e}/\mathcal{M}}  \label{del}
\end{equation}%
which for a sufficiently high mass scale $\mathcal{M}$ happens by itself (as
we see below) to be of a certain interest for current high-energy tests of
special relativity. Particularly, when taking just the Planck mass $M_{P}$
for the highest scale in the theory ($\mathcal{M}=M_{P}$), one has the upper
limits 
\begin{equation}
\left\vert \delta \right\vert \lesssim \overline{\delta }=6.5\times 10^{-12},%
\text{ \ \ }M\lesssim 10^{8}(e/\Delta e)\text{ }GeV  \label{del1}
\end{equation}%
for the $\delta $ parameter and the Lorentz violation scale $M$,
respectively. \ 

Before proceeding to applications, let us note that, in the order $1/%
\mathcal{M}$ considered, all other particles apart from charged fermions,
such as photon, neutrinos, weak bosons etc. are proposed to satisfy the
standard dispersion relations. Inclusion of new charged fermions into the
Goldstonic QED Lagrangian (\ref{NL}) will in general increase the number of
the SLIV$\ $ parameters in theory by assigning to every fermion species $f$
(being some lepton or quark) its own $\delta _{f}$ parameter . These
parameters, as is seen from (\ref{del}), will actually differ from one
another by the corresponding charge differences $(\Delta e)_{f}$ only. This
immediately leads to the conclusion that the $\delta $ parameters for
particles and antiparticles must be equal but of opposite sign. Apart from
that, some of the charge differences might appear to be equal if certain
symmetries for leptons and quarks are postulated; say, grand unified
symmetry inside a lepton-quark family and/or flavor symmetry between
families.

\section{ Some immediate applications}

One may now see that, due to spontaneous Lorentz violation resulting in the
new dispersion relation (\ref{dr}) for charged fermions, the kinematics of
processes in which such fermions are participating is substantially changed.
At low energies these changes can be neglected, but at high energies they
may play crucial role. As a result, some of allowed processes appear to be
suppressed at high energies and, on the contrary, some of suppressed
processes are now allowed to go. This could substantially change the
particle phenomenology at high energies that would lead to some new
observations, as well as corrections to the early Universe scenario. Certain
of these processes were previously discussed in direct Lorentz violation
scenarios [2,3]. Predictions of our SLIV model appear in fact to be more
distinctive being dependent on only a few SLIV parameters $\delta $ (\ref%
{del}) assigned to elementary charged fermions, quarks and leptons.
Actually, all changes as compared with a conventional QED can readily be
derived replacing masses $m_{f}$\ of these fermions by their non-covariant
"effective" masses%
\begin{equation}
m_{f}^{\ast }\equiv \sqrt{p_{\mu }^{2}}\cong \left\vert m_{f}+2\boldsymbol{%
\delta }_{f}p_{0}\right\vert ,  \label{eff}
\end{equation}%
as follows from the above dispersion relation (\ref{dr}), where we also
introduced a modified (two-component) parameter $\boldsymbol{\delta }_{f}$
which is equal to $\boldsymbol{\delta }_{f}=\delta _{f}$ for the time-like
SLIV and $\boldsymbol{\delta }_{f}=\delta _{f}\cos \theta $ for the
space-like SLIV, respectively. Note that in a high-energy region that we are
interested in, the scalar product $p_{\mu }n^{\mu }/n^{2}$ in Eq.(\ref{dr})
for the space-like SLIV ($n^{2}<0$) with the angle $\theta $ between a
fermion 3-momentum $\overrightarrow{p}$ and the Lorentz violation direction
vector $\overrightarrow{n}$ just reduces to $p_{\mu }n^{\mu
}/n^{2}=\left\vert \overrightarrow{p}\right\vert \cos \theta $ $\cong $ $%
p_{0}\cos \theta $.

Consideration of composite hadron states, mesons and baryons, in our model
needs further clarification. Generally, one could assign to each of these
composites its own $\boldsymbol{\delta }$ parameter, or its own effective
mass $m^{\ast }$ (\ref{eff}) which would lead to a plethora of new SLIV
parameters in the model. However, we propose the following simple rules for
composites that might naturally work. Actually, one may treat SLIV features
of hadrons solely based on their quark content so that their effective
masses are additively combined with those of quarks and antiquarks involved,
both taken at the same energy $E$ in a preferred frame. So, for some meson $%
\varphi $ consisting of quark $q_{1}$ and antiquark $\overline{q}_{2}$ this
effective mass might look like%
\begin{equation}
m_{\varphi }^{\ast }=\left\vert m_{\varphi }+2(\boldsymbol{\delta }_{1}-%
\boldsymbol{\delta }_{2})E\right\vert  \label{12}
\end{equation}%
where we have used that the $\boldsymbol{\delta }$ parameter for antiquark
has an opposite sign ($\boldsymbol{\delta }_{q_{1}}\equiv \boldsymbol{\delta 
}_{1}$, $\boldsymbol{\delta }_{\overline{q}_{2}}=-\boldsymbol{\delta }%
_{q_{2}}\equiv -\boldsymbol{\delta }_{2})$, as was indicated at the end of
Sec.2, and replaced the sum of the current quark masses $m_{1}+m_{2}$ in (%
\ref{12}) by the meson invariant mass $m_{\varphi }$. This replacement seems
to be a quite good approximation for mesons consisting of heavy $c$, $b$ and 
$t$ quarks, but not for mesons consisting of light quarks $u$, $d$ and $s$,
whose current masses $m_{u,d,s}$ hardly provide masses of the corresponding
mesons (and baryons). The point is, however, that the color interaction
converting these current quark masses into the constituent quark ones (and
leading eventually to the physical hadron masses) is presumably Lorentz
invariant, so that the non-covariant part in \ the meson effective mass (\ref%
{12}) with $\boldsymbol{\delta }$ parameters depending solely on the quark
electric charge differences $(\Delta e)_{q_{1,2}}$ seems to be basically
preserved. Analogously, dispersion relations of baryons are always
frame-dependent being determined by a particular quark content in its
effective mass 
\begin{equation}
m_{B}^{\ast }=\left\vert m_{B}+2(\boldsymbol{\delta }_{1}+\boldsymbol{\delta 
}_{2}+\boldsymbol{\delta }_{3})E\right\vert  \label{123}
\end{equation}%
provided that the baryon $B$ with invariant mass $m_{B}$ is composed of
quarks $q_{1}$, $q_{2}$ and $q_{3}$ with parameters $\boldsymbol{\delta }%
_{1,2,3}$.

A few simple remarks are in order. As is readily seen from Eq.(\ref{12}),
mesons which are diagonal in the quark flavors, such as $\pi ^{0}$, $\eta ,$ 
$\rho ^{0},$ $\phi ,$ $J/\Psi $ etc., have zero $\boldsymbol{\delta }$
parameters and thus they hold standard dispersion relations. Furthermore,
mesons and baryons with the same quark content possess equal $\boldsymbol{%
\delta }$ parameters and, therefore, have alike effective masses. And, in a
similar manner as in the elementary fermion case, the $\boldsymbol{\delta }$
parameters for composite hadrons and their antiparticles appear to be equal
but of opposite sign.

\subsection{GZK cutoff revised}

One of the most interesting examples where a departure from Lorentz
invariance can essentially affect a physical process is the transition $%
p+\gamma \rightarrow \Delta $ which underlies the Greisen-Zatsepin-Kouzmin
(GZK) cutoff for ultra-high energy (UHE) cosmic rays \cite{gzk}. According
to this idea primary high-energy nucleons ($p$ ) should suffer an inelastic
impact with cosmic background photons ($\gamma $) due to the resonant
formation of \ the first pion-nucleon resonance $\Delta (1232)$, so that
nucleons with energies above $\sim 5\cdot 10^{19}eV$ could not reach us from
further away than $\sim 50$ $Mpc$. During the last decade there were serious
indications \cite{GZK} that the primary cosmic-ray spectrum extends well
beyond the GZK cutoff, though presently the situation is somewhat unclear
due to a certain criticism of these results and new data that recently
appeared \cite{Oge}. However, no matter how things will develop, we could
say that according to the new fermion dispersion relation (\ref{dr}) the GZK
cutoff will necessarily be changed (increased or decreased depending on the
sign of the corresponding $\delta $ parameter) at superhigh energies.
Remarkably, for the Planck mass scale case in the theory ($\mathcal{M}$ $%
=M_{P}$) the above transition, providing this cutoff, appears to be
significantly weakened (or even completely undone) just around the
aforementioned GZK energy region, as one can see from the $\delta $
parameter value range (\ref{del1}) calculated for this case.

Really, we must replace the fermion masses in a conventional proton
threshold energy for this process by their effective masses $m_{p,\Delta
}^{\ast }\cong $ $\left\vert m_{p,\Delta }+2\boldsymbol{\delta }E_{p,\Delta
}\right\vert $ which can be taken with equal $\boldsymbol{\delta }$
parameters as for composite states having a similar quark content ($%
\boldsymbol{\delta }\equiv \boldsymbol{\delta }_{p}=\boldsymbol{\delta }%
_{\Delta }=2\boldsymbol{\delta }_{u}+\boldsymbol{\delta }_{d}$). Using then
the approximate equality of their energies, $E_{\Delta }=E_{p}+\omega \cong
E_{p}$, since the target photon energies $\omega $ are vanishingly small
(being a thermal distribution with temperature $\ T=2.73$ $K$, or $kT\equiv 
\overline{\omega }=2.35\times 10^{-4}eV$), one comes to the condition
determining the proton energy region in which the foregoing transition is
kinematically forbidden for a head-on impact 
\begin{equation}
E_{p}>\frac{m_{\Delta }^{2}-m_{p}^{2}}{4[\omega -\boldsymbol{\delta }%
(m_{\Delta }-m_{p})]}=\frac{6.8}{\omega /\overline{\omega }-8.1\text{ }%
\boldsymbol{\delta }\mathbf{/}\overline{\delta }}\times 10^{20}\text{ }eV.
\label{thre}
\end{equation}%
As one can readily see, the SLIV modification of the proton threshold energy 
$E_{p}$ in Eq.(\ref{thre}) might naturally relax the GZK cutoff and even
permits UHE cosmic-ray nucleons to travel cosmological distances (when $%
\omega /\overline{\omega }\approx 8.1$ $\boldsymbol{\delta }\mathbf{/}%
\overline{\delta }$ with $\overline{\delta }$ given in (\ref{del1}))
provided that the $\boldsymbol{\delta }$ parameter in Eq.(\ref{thre}) is
taken positive. Conversely, for its negative values the original GZK cutoff
tends to a decrease. Most interestingly, there is predicted some marked
spatial anisotropy for primary nucleons in the space-like Lorentz violation
case ($\boldsymbol{\delta }$ $=$ $\delta \cos \theta $) which results in an
ordinary GZK cutoff for perpendicular (to the SLIV vector $\overrightarrow{n}
$) direction, whereas it is lower or higher for other directions.

\subsection{Stability of high-energy vector and scalar bosons}

Another interesting example is provided by decays of vector and scalar
bosons into\ fermions, no matter whether they all are elementary or
composite. Usually these processes are possible if a boson mass $\mathfrak{m}
$ is no less than the sum of fermion invariant masses $m_{1,2}$, but now,
when fermions and (composite) bosons can have some effective masses given by
Eqs.(11,12,13), these decays at high energies may appear to be kinematically
suppressed, as can easily be confirmed. Actually, for a particular two-body
decay case this process appears to be banned if the inequality $\mathfrak{m}%
^{\ast }<m_{1}^{\ast }+m_{2}^{\ast }$ for fermion effective masses $%
m_{1,2}^{\ast }$ is satisfied for the minimum total energy of decay products
with a given total momentum $\overrightarrow{P}$. It follows that all
momenta are collinear \ in the configuration of minimum total energy and
fermion momenta are equal to 
\begin{equation}
\overrightarrow{p}_{1,2}=\frac{m_{1,2}}{m_{1}+m_{2}}\overrightarrow{P}
\end{equation}%
so that at energies%
\begin{equation}
E>\frac{1}{2}(\mathfrak{m}-m_{1}-m_{2})\frac{m_{1}+m_{2}}{\boldsymbol{\delta 
}_{1}m_{1}+m_{2}\boldsymbol{\delta }_{2}-\boldsymbol{\delta }(m_{1}+m_{2})}
\label{eb}
\end{equation}%
this boson could appear stable.

Applying this result to the weak $W$ boson decays into quarks and leptons ($%
\mathfrak{m}=m_{W},$ $\boldsymbol{\delta }\equiv \boldsymbol{\delta }_{W}=0$)%
\footnote{%
For the pure leptonic decay $W\rightarrow l\overline{\nu }$ the equation (%
\ref{eb}) is maximally simplified, $E_{B}>(m_{W}-m_{l})/2\boldsymbol{\delta }%
_{l}$, since neutrino is presumably massless and has a normal dispersion
relation ($m_{2}=0$ and $\delta _{2}=0$).} and taking the $\boldsymbol{%
\delta }_{1,2}$ parameters to be of the same order as those that are
required for a weakened GZK cutoff version ($\boldsymbol{\delta }_{1,2}\sim 
\boldsymbol{\delta }_{p,\Delta }\sim 10^{-12}$), we find \ that stable $W$
bosons appear at the energy region $\sim 10^{23}eV$ that seems to be
somewhat problematic to be directly detected. At the same time, the $Z$ and
Higgs bosons which are only related to the flavor-diagonal quark and lepton
currents do not change their decay rates with energy since, as already
noted, the SLIV effects from particles and antiparticles are expected to be
cancelled.

However, the special observational interest may cause charged pion stability
at high energies against the standard $\pi \rightarrow \mu +\nu $ decays. In
contrast to the $W$ boson, the composite charged pion has a non-zero SLIV
parameter $\boldsymbol{\delta }_{\pi }=\boldsymbol{\delta }_{u}-\boldsymbol{%
\delta }_{d}$ (expressed in the up and down quark parameters $\boldsymbol{%
\delta }_{u,d}$, see Eq.(\ref{12})) and, therefore, the non-covariant
effective mass $m_{\pi }^{\ast }=\left\vert m_{\pi }+2\boldsymbol{\delta }%
_{\pi }E_{\pi }\right\vert $. So, properly adjusting a general formula (\ref%
{eb}) for two-body decays ($\mathfrak{m}=m_{\pi },$ $\boldsymbol{\delta }=%
\boldsymbol{\delta }_{\pi };$ $\ m_{1}=m_{\mu },$ $\boldsymbol{\delta }_{1}=%
\boldsymbol{\delta }_{\mu };$ $m_{2}=m_{\nu }=0,$ $\boldsymbol{\delta }_{2}=%
\boldsymbol{\delta }_{\nu }=0$) one eventually has for the threshold energy
providing pion stability ($m_{\pi }^{\ast }<m_{\mu }^{\ast }$) 
\begin{equation}
E_{\pi }>\frac{1}{2}\frac{m_{\pi }-m_{\mu }}{\boldsymbol{\delta }_{\mu }-%
\boldsymbol{\delta }_{\pi }}
\end{equation}%
This energy region, when the muon and pion $\boldsymbol{\delta }$ parameter
values ($\boldsymbol{\delta }_{\mu }-\boldsymbol{\delta }_{\pi }>0$) are
taken to be of the same order $\sim 10^{-12}$ as in the foregoing cases,
appears to be significantly lower than that for the stable $W$ boson, being
just near the GZK cutoff energy $\sim 10^{19}eV$. Thus, the UHE primary
cosmic rays may include stable charged pions that could in principle be
detected at current experiments\cite{Oge}, whereas neutral pions being
diagonal quark-antiquark composites are left to be very unstable, as they
usually are. Again, for the space-like SLIV case the spatial anisotropy is
expected according to which the stable charged pions are predicted to be
largely located along the SLIV direction $\overrightarrow{n}$.

\subsection{Modified nucleon decays}

As a last example we consider an ordinary neutron $\beta $ decay ($%
n\rightarrow pe^{-}\overline{\nu }$). Since the neutron is heavier than the
proton, $m_{n}>m_{p}$, usually neutron $\beta $ decay is allowed, while
proton $\beta $ decay ($p\rightarrow ne^{+}\nu $) is kinematically
suppressed. However, due to the Lorentz non-invariance their effective
masses may grow at high energies in such a way that $m_{n}^{\ast
}<m_{p}^{\ast }$ in a preferred frame. This means that neutrons and protons
change places - neutrons become stable, whereas protons decay. Using the
above general formula (\ref{eb}) one can readily find the threshold energy
value when this happens, 
\begin{equation}
E>\frac{m_{n}-m_{p}}{2(\boldsymbol{\delta }_{p}-\boldsymbol{\delta }_{n})}=%
\frac{m_{n}-m_{p}}{2(\boldsymbol{\delta }_{u}-\boldsymbol{\delta }_{d})}%
\text{ ,}
\end{equation}%
where we have treated both beta processes as essentially two-body decays
with lepton masses ignored. Again with the $\boldsymbol{\delta }_{p,n}$
parameters $\sim 10^{-12}$ taken as in the foregoing examples, one finds to
the energy region $E>10^{18}eV$ \ which is an active research area for
current cosmic-ray experiments [14,15]. At these energies stable neutrons,
as follows, can be contained in primary UHE cosmic rays, whereas unstable
protons cannot.

To conclude, we have considered some basic applications of the model which
happen to be described in terms of a few $\boldsymbol{\delta }$ parameters
assigned to elementary fermions, quarks and leptons. Our Lorentz violating
predictions appear to be quite certain for the above processes being
conditioned just by the vector field model of the SLIV. At the same time
this minimal model predicts the strictly vanishing effects in many processes
(where generally some Lorentz violation might in principle be expected),
such as the $Z$ and Higgs boson and photon decays, decays of diagonal
quark-antiquark composites ($\pi ^{0}$, $\eta ,$ $\rho ^{0},$ $\phi ,$ $%
J/\Psi $ etc.), neutrino oscillations and others which have been previously
discussed on pure hypothetical grounds [2,3].

\section{Summary and outlook}

We have argued that genuine spontaneous Lorentz violation, which in the QED
framework would induce the photon as a vector NG boson, does not manifest
itself in any physical way due to the gauge invariance involved. In
substance, the SLIV ansatz taken as $A_{\mu }(x)=a_{\mu }(x)+n_{\mu }M$ \
may be treated by itself as a pure gauge transformation with gauge function
linear in coordinates, $\omega (x)=$ $(n_{\mu }x^{\mu })M$. In this sense,
gauge invariance in the QED \ leads to the conversion of the SLIV into gauge
degrees of freedom of the massless photon. This is what one could refer to
as the generic non-observability of the SLIV in the QED. Furthermore, as was
shown some time ago \cite{cfn}, gauge theories, both Abelian and
non-Abelian, can be obtained by themselves from the requirement of the
physical non-observability of the SLIV, caused by the Goldstonic nature of
vector fields, rather than from the standard gauge principle.

All this requires that gauge invariance in the QED should be broken rather
than exact. We have proposed some simple model for a tiny gauge
non-invariance that might be caused by quantum gravity at extra-small
distances through some higher order operators involved. To this end we
extended QED to the lowest order in $1/\mathcal{M}$ (the theory's inverse
scale) so that all possible dimension-five operators compatible with
accompanying global and discrete symmetries are included. They appear to
possess in general some approximate gauge invariance rather than an exact
one as in a conventional QED with dimensionless coupling constants. This in
turn leads of necessity to the physical Lorentz violation resulting in the
new dispersion relation (\ref{dr}) for charged fermions. As a consequence,
kinematics of processes in which such fermions are participating is
substantially changed. While at low energies these changes can be neglected,
at high energies they may play crucial role, as was illustrated by the
examples of processes given above.

We have so far considered the QED theory by itself. However, this theory
should be included into the Standard Model with its internal gauge symmetry $%
SU(2)\times U(1)_{Y}$ to have fully realistic framework. This will
substantially change an entire approach, though the physical consequences
are largely preserved and, moreover, extended. Note first of all that the
SLIV condition (1) is taken now for the $U(1)_{Y}$ hypercharge gauge field $%
B_{\mu }$ rather than for the electromagnetic one $A_{\mu }$ which by itself
appears later when the starting $SU(2)\times U(1)_{Y}$ symmetry is
spontaneously broken. Furthermore, for the physical Lorentz violation to
occur this $U(1)_{Y}$ hypercharge gauge invariance should be explicitly
broken by some high-dimension operators supposedly induced by quantum
gravity. It is apparent, on the other hand, that the chiral nature of this
symmetry in the SM forbids quarks and leptons to form the "bosonic" type
terms which in the pure QED framework were taken for a vectorlike fermion in
the Lagrangians (\ref{free}, \ref{lagr}). This means that there cannot
appear the $U(1)_{Y}$ breaking terms in $1/\mathcal{M}$ order, as happened
in the above QED case. However, they appear, as one can readily see, in the
next order $1/\mathcal{M}^{2}$. Using again the restriction requirements
related to all discrete and global symmetries involved one can reduce a
number of possible breaking terms to a few ones including some "gravity
type" coupling 
\begin{equation}
\frac{1}{\mathcal{M}^{2}}B_{\mu }B_{\nu }\Theta ^{\mu \nu }  \label{fi}
\end{equation}%
where $\Theta ^{\mu \nu }$ stands for a total energy-momentum tensor of all
fermion and vector fields involved, which is proposed to be symmetrical,
conserving and gauge invariant\footnote{%
Among possible couplings the foregoing breaking terms emerging in the pure
QED Lagrangian (\ref{lagr}) could of course be induced in the SM. They arise
when dimension-six coupling $(\phi /\mathcal{M}^{2})D_{\mu }^{\prime \ast }%
\overline{\psi }_{L}D^{\prime \mu }\psi _{R}$ appears and the conventional
Higgs doublet $\phi $ acquires its vev. However, these terms are down by the
tiny ratio of the electroweak scale to the SLIV one as compared with the
chirality preserving couplings.}. Remarkably enough, the coupling (\ref{fi})
appears as the only possible one, if one further requires that these
breaking terms possess a partial $U(1)_{Y}$ gauge symmetry in the sense
that, while the theory is basically $U(1)_{Y}$ gauge invariant (being
constructed from ordinary covariant derivatives of all matter fields
involved), the $B_{\mu }$ field by itself is allowed to form its own
polynomial couplings\footnote{%
These couplings are like the potential terms (\ref{pol}) introduced in the
QED case - they in a similar manner will lead to the SLIV condition $%
B^{2}=n^{2}M^{2}$.} and it may also appear as factors in other field
couplings. Such couplings, and specifically the coupling (\ref{fi}), cause a
tiny $U(1)_{Y}$ gauge non-invariance in the SM and lead eventually to the
physical Lorentz violation. As a result, after a spontaneous $SU(2)\times
U(1)_{Y}$ symmetry breakdown, one comes to the changed dispersion relation
for all quarks and leptons including neutrinos, since they all possess
hypercharge and thus are related to the vector field $B_{\mu }$ developing a
vev. Another peculiarity related to an extension to the SM is that the gauge
non-invariance now appears not only in the fermion sector of theory, but
also in the vector field sector in itself. So, one eventually comes to the
changed dispersion relation for the photon as well that leads in turn to
extra interesting manifestations to be observed. The entire framework for a
study of spontaneous Lorentz violation in the Standard Model with non-exact
gauge invariance is planned to be considered in detail elsewhere.

\section*{Acknowledgments}

One of us (J.L.C.) thanks J. Bjorken, C. Froggatt, R. Jackiw, R. Mohapatra
and H. Nielsen for interesting correspondence, useful discussions and
comments. Financial support from World Federation of Scientists is
gratefully acknowledged by J.L.C. and Z.K.

\end{document}